\documentclass[%
twocolumn,
preprintnumbers,
 amsmath,amssymb,
 aps,
 american
 prd,
 superscriptaddress,
]{revtex4-2}
\usepackage[paperwidth=210mm,paperheight=297mm,centering,hmargin=1.25cm,vmargin=2.0cm]{geometry}
\usepackage{amsmath,bm,dsfont,slashed,ulem,color,float,lineno}
\usepackage[english]{babel}
\usepackage{graphicx,dcolumn,hyperref}
\begin{document}

\title{Exploring Anisotropic Effects in Magnetized Quark Matter}
\author{S.A. Ferraris}
\email{sebastianferraris@cnea.gob.ar}
\affiliation{Departamento de F\'\i sica, Comisi\'on Nacional de Energ\'{\i}a At\'omica, Avenida del Libertador 8250, (1429) Buenos Aires, Argentina}

\author{J.P. Carlomagno}
\affiliation{IFLP, CONICET departamento de Física, Facultad de Ciencias Exactas, Universidad Nacional de La PLata, C.C. 67, (1900) La Plata, Argentina}
\affiliation{CONICET, Godoy Cruz 2290, Buenos Aires, Argentina} 

\author{G.A. Contrera}
\email{contrera@fisica.unlp.edu.ar}
\affiliation{IFLP, CONICET departamento de Física, Facultad de Ciencias Exactas, Universidad Nacional de La PLata, C.C. 67, (1900) La Plata, Argentina}
\affiliation{CONICET, Godoy Cruz 2290, Buenos Aires, Argentina}

\author{A.G. Grunfeld}
\email{ag.grunfeld@conicet.gov.ar}
\affiliation{Departamento de F\'\i sica, Comisi\'on Nacional de Energ\'{\i}a At\'omica, Avenida del Libertador 8250, (1429) Buenos Aires, Argentina}
\affiliation{CONICET, Godoy Cruz 2290, Buenos Aires, Argentina}

\date{\today}

\begin{abstract}
We investigate the thermodynamic properties of cold magnetized quark matter within a nonlocal Nambu–Jona-Lasinio (nlNJL) model. Our study addresses the equation of state, anisotropic pressures, quark density, speed of sound, and magnetic susceptibility, with direct comparison to the chiral limit. Strong magnetic fields are found to generate marked anisotropy: the longitudinal pressure and speed of sound are enhanced, approaching the causal bound in the lowest Landau-level (LLL) regime, while the transverse components are systematically reduced. The quark density exhibits magnetic catalysis, increasing with both the chemical potential and the magnetic field strength. 
At moderate to high fields, the critical chemical potential decreases with increasing $eB$, signaling the occurrence of inverse magnetic catalysis at finite chemical potential ($\mu$IMC).
Magnetic susceptibility displays oscillations around zero in low fields, driven by de Haas–van Alphen–like effects, and settles at positive values for strong fields, consistent with an overall growth of magnetization. Compared with the chiral limit, the inclusion of finite current quark masses does not modify the overall oscillatory behavior, but changes the nature of the Landau-level transitions, which become weakly first order instead of second order.
\end{abstract}


\maketitle

\section{Introduction}

The study of pressure anisotropy in magnetized quark matter provides fundamental insight into how strong magnetic fields modify the thermodynamic response of strongly interacting systems. A uniform magnetic field explicitly breaks spatial isotropy, leading to different pressures along and perpendicular to the field direction. The longitudinal pressure, $P_{\parallel}$, and the transverse one, $P_{\perp}$, thus carry distinct information about the coupling between the magnetic field and the microscopic degrees of freedom of the medium. This anisotropy is not a secondary effect: it has a direct impact on the equation of state (EOS), transport properties, and dynamical behavior of dense matter~\cite{Ferrer:2010wz,Paulucci:2010uj,Andersen:2014xxa,Miransky:2015ava}.

Strong magnetic fields are known to appear in various physical contexts~\cite{Adhikari:2024bfa}. In astrophysics, magnetars are expected to exhibit surface fields up to $\sim 10^{15}\,\mathrm{G}$, with possibly higher values in their cores, where baryon densities exceed nuclear saturation and deconfined quark matter may exist~\cite{Duncan:1992hi,Thompson:1995gw}. Under such extreme conditions, the anisotropy between $P_{\parallel}$ and $P_{\perp}$ can lead to structural deformations, influence the stability of compact stars, and modify their global properties~\cite{Chatterjee:2018prm}. The magnetic field also affects the EOS and the propagation of sound within the stellar core, with implications for mass–radius relations, tidal deformabilities, and thermal evolution.

Similar effects are expected in relativistic heavy-ion collisions, where noncentral impacts generate transient magnetic fields reaching up to $\sim 10^{18}\,\mathrm{G}$~\cite{Skokov:2009qp,Tuchin:2013apa}. These fields introduce anisotropic features in the quark–gluon plasma (QGP), altering its hydrodynamic expansion and collective flow. The resulting differences between $P_{\parallel}$ and $P_{\perp}$ can leave measurable imprints on observables such as the elliptic flow coefficient $v_2$, offering an indirect way to probe the magnetic response of strongly interacting matter.

From a microscopic perspective, the pressure anisotropy reflects the magnetization of the system, which measures its response to an external magnetic field. The difference between the longitudinal and transverse pressures is intimately connected to the magnetic energy stored in the medium and to phenomena such as Landau-level quantization, magnetic catalysis, and effective dimensional reduction in the strong-field regime~\cite{Kharzeev:2012ph,Andersen:2014xxa,Ferrer:2015wca}. When the system becomes dominated by the lowest Landau level (LLL), the transverse dynamics are largely frozen, leading to a stiffening of the longitudinal sector and a pronounced anisotropy in the EOS.

Since first-principles lattice QCD calculations remain limited at finite baryon density, effective chiral models constitute an essential framework to explore the thermodynamic and magnetic properties of dense quark matter. The Nambu–Jona-Lasinio (NJL) model~\cite{Nambu:1961tp,Nambu:1961fr,Vogl:1991qt,Klevansky:1992qe,Hatsuda:1994pi,Buballa:2003qv,Menezes:2008qt,Costa:2013zca,Andersen:2014xxa} and its nonlocal extensions~\cite{Contrera:2007wu,Contrera:2010kz,Ferreira:2013tba,Pagura:2016pwr,GomezDumm:2017iex,Dumm:2021vop} have proven particularly useful for describing the interplay between chiral symmetry breaking and external magnetic fields. Nonlocal formulations, in particular, yield improved agreement with hadronic observables and lattice QCD results, enabling a realistic description of magnetized quark matter at finite chemical potential. 
A noteworthy feature observed in various effective approaches, including the present nonlocal framework, is that for moderately strong magnetic fields the critical chemical potential for chiral symmetry restoration tends to decrease as $eB$ increases. This behavior, opposite to the usual magnetic catalysis, is commonly referred to as “inverse magnetic catalysis at finite chemical potential” ($\mu$IMC) ~\cite{Costa:2015bza,Pagura:2016pwr}. Similar effects have been reported in local NJL-type models~\cite{Carlomagno:2023clk} and in holographic approaches such as the Sakai–Sugimoto model~\cite{Preis:2010cq,Preis:2012fh}.

Within this context, analyzing the anisotropy between $P_{\parallel}$ and $P_{\perp}$ provides a powerful diagnostic of how magnetic fields influence the internal dynamics of dense matter. It also offers a link between microscopic QCD mechanisms and macroscopic observables relevant to astrophysics and heavy-ion physics. 

In this work, we investigate anisotropic effects in cold and dense quark matter within a covariant nonlocal two-flavor Nambu–Jona-Lasinio (nlNJL) model. Our analysis focuses on the dependence of the anisotropic pressures, magnetization, magnetic susceptibility, quark number density, and sound velocities on both the magnetic field strength and chemical potential, comparing the chiral limit with the case of finite current quark masses. The article is organized as follows. In Sec.~\ref{sect2} we present the theoretical formalism for magnetized quark matter within the nlNJL model. Sect.~\ref{sect3} contains our numerical results, and in Sec.~\ref{sect4} we summarize our main conclusions.

\section{Formalism}
\label{sect2}

We introduce the Euclidean action of the nonlocal Nambu--Jona-Lasinio (nlNJL) model for two light quark flavors, $u$ and $d$:

\begin{equation}
S_{E} = \int d^4x \left[\bar{\psi}(x)\left(-i\slashed{\partial} + m_c\right)\psi(x) - \frac{G}{2} j_a(x) j_a(x) \right] \ , \label{accionE}
\end{equation}
where $\psi$ denotes the quark fields, $G$ is the scalar coupling constant, and $m_c$ represents the common current mass for both flavors, i.e., $m_c = m_u = m_d$. Since our aim is to analyze the response of quark matter to an external magnetic field, the pure electromagnetic field term $F_{\mu\nu}F^{\mu\nu}/4$ is not included in the effective action. The nonlocal nature of the interaction is encoded in the bilinear quark currents, which are defined as
\begin{equation}
j_a(x) = \int d^4z\ \mathcal{G}(z)\, \bar{\psi}\left(x + \frac{z}{2}\right) \Gamma_a \psi\left(x - \frac{z}{2}\right) \ , \label{corrientes}
\end{equation}
with $\Gamma_a = (\mathds{1},\, i\gamma_5 \vec{\tau})$, and $\mathcal{G}(z)$ a nonlocal form factor that modulates the interaction range.

To include the coupling to an external electromagnetic field $\mathcal{A}_\mu$, the derivative in the kinetic term must be promoted to a covariant derivative,
\begin{equation}
\partial_\mu \rightarrow D_\mu \equiv \partial_\mu - i \hat{Q} \mathcal{A}_\mu \ ,
\end{equation}
where $\hat{Q} = \text{diag}(q_u, q_d)$ is the flavor charge matrix, with $q_u = 2e/3$ and $q_d = -e/3$. This substitution modifies the kinetic term accordingly and also requires a gauge-invariant extension of the nonlocal currents in Eq.~(\ref{corrientes})~\cite{Schmidt:1994di,GomezDumm:2006vz,Noguera:2008cm,GomezDumm:2010cta}. Following the standard approach, gauge invariance is maintained by introducing a straight-line path between the points $x \pm z/2$ in the nonlocal term.

For definiteness, we consider a constant and homogeneous magnetic field $\vec{B}$ pointing along the $x_3$ axis. We choose the Landau gauge, $\mathcal{A}_\mu = B x_1 \delta_{\mu 2}$, which satisfies $\vec{\nabla} \times \vec{\mathcal{A}} = \vec{B}$. 

Since quark fields are not physical degrees of freedom in the infrared, it is convenient to integrate them out and rewrite the theory in terms of auxiliary bosonic fields. This bosonization step,~\cite{Noguera:2008cm,GomezDumm:2010cta,GomezDumm:2017jij}, introduces the fields of the scalar and pseudoscalar mesons $\sigma(x)$ and $\vec{\pi}(x)$, respectively.

In what follows, we adopt the mean field approximation (MFA), taking the scalar field to acquire a nontrivial constant value $\bar{\sigma}$, while the mean values of the pseudoscalar fields are taken to vanish, $\pi_a(x) = 0$, due to flavor symmetry considerations.

Since our goal is to explore the properties of dense quark matter, we work at finite quark chemical potential $\mu$, which relates to the baryon chemical potential via $\mu = \mu_B/3$. The grand canonical thermodynamic potential can be derived from the effective action by incorporating $\mu$ through the standard replacement $\partial_4 \rightarrow \partial_4 - i \mu$ in the kinetic term. Furthermore, to ensure that conserved currents are correctly described, one must also adapt the nonlocal currents introduced in Eq.~(\ref{corrientes})~\cite{Ferraris:2021vun}. In particular, if $g(p)$ denotes the Fourier transform of the nonlocal form factor $\mathcal{G}(z)$, the chemical potential is included by shifting $p_4 \rightarrow p_4 - i\mu$.
Following this prescription, the mean-field thermodynamic potential in the presence of a magnetic field reads
\begin{eqnarray}
\Omega_{\mu,B}^{\rm MFA} &=& \frac{\bar{\sigma}^{2}}{2G} -
\sum_{f=u,d} \frac{3 B_f}{2\pi} \int \frac{d^{2}p_\parallel}{(2\pi)^{2}} \nonumber\\
&\times& \left[\ln\left(p_\parallel^{2} + {M_{0,p_\parallel}^{s_f,f}}^{\,2\,}\right)
+ \sum_{k=1}^{\infty} \ln \Delta_{k,p_\parallel}^{f} \right] \ , \label{gpotencialB}
\end{eqnarray}
where
\begin{eqnarray}
\Delta_{k,p_\parallel}^{f} &=& \left(2k B_f + p_\parallel^{2} +
M_{k,p_\parallel}^{+,f} M_{k,p_\parallel}^{-,f}\right)^{2} \nonumber\\
&& +\, p_\parallel^{2} \left(M_{k,p_\parallel}^{+,f} - M_{k,p_\parallel}^{-,f}\right)^{2} \ , \label{deltaKP}
\end{eqnarray}
with the effective masses
\begin{equation}
M_{k,p_\parallel}^{\lambda,f} = \left(1 - \delta_{k_\lambda,-1}\right) m_c
+ \bar{\sigma}\ g_{k,p_\parallel}^{\lambda,f}
\label{const_mass}
\end{equation}
and
\begin{eqnarray}
g_{k,p_\parallel}^{\lambda,f} &=&
\frac{4\pi}{|q_f B|} (-1)^{k_\lambda} \int \frac{d^{2}p_\bot}{(2\pi)^2}
g\left(p_\bot^2 + p_\parallel^2\right) \nonumber \\
&& \times \exp\left(-\frac{p_\bot^2}{B_f}\right)
L_{k_\lambda}\left(\frac{2p_\bot^2}{B_f}\right)\ . \label{funcg}
\end{eqnarray}
In the above expressions we adopt the conventions $p_{\bot}=(p_1,p_2)$, and $p_\parallel=(p_3,p_4-i\mu)$. The Landau-level index is given by $k_{\pm} = k - 1/2 \pm s_f/2$, with $s_f = \text{sign}(q_f B)$, and $B_f = |q_f B|$. The functions $L_m(x)$ denote the generalized Laguerre polynomials, where we use the standard convention $L_{-1}(x) = 0$. The quantities $M_{k,p_\parallel}^{\pm,f}$ can be interpreted as constituent quark masses modified by the magnetic field, while $k$ indexes the Landau levels arising from the quantization of the transverse momentum.

The expression in Eq.~(\ref{gpotencialB}) is ultraviolet divergent and requires regularization. Following the method proposed in Ref.~\cite{GomezDumm:2004sr}, the regularized thermodynamic potential is written as
\begin{equation}
\Omega_{\mu,B}^{\rm MFA,reg} = \Omega_{\mu,B}^{\rm MFA} -
\Omega_{\mu,B}^{\rm free} + \Omega_{\mu,B}^{\rm free,reg} \ , \label{omereg}
\end{equation}
where $\Omega_{\mu,B}^{\rm free}$ is the free (noninteracting) contribution obtained from $\Omega_{\mu,B}^{\rm MFA}$ by setting $\bar{\sigma} = 0$, while keeping $\mu$ and $B$ fixed. Its regularized form reads
\begin{eqnarray}
\Omega_{\mu,B}^{\rm free,reg} &=& -\frac{N_c}{2\pi^2} \sum_{f=u,d}
\Big[ B_f^2\, t_f \nonumber \\
&& + \sum_k \theta\left(\mu - S_{kf}\right)\, \alpha_k\, B_f\, v_{kf} \Big] \ , \label{omegafree}
\end{eqnarray}
with the auxiliary functions
\begin{equation}
t_f = \zeta'\left(-1,x_f\right) + \frac{x_f^2}{4} - \frac{1}{2}(x_f^2 - x_f) \ln x_f \ ,
\end{equation}
\begin{equation}
v_{kf} = \mu \sqrt{\mu^2 - S_{kf}^2} - S_{kf}^2
\ln\left[\frac{\mu + \sqrt{\mu^2 - S_{kf}^2}}{S_{kf}}\right] \ .
\end{equation}
Here we have used the definition 
$\zeta^{'}(-1,x_{f})=d\zeta(z,x_{f})/dz|_{z=-1}$.
Finally, the mean-field value $\bar{\sigma}$ is obtained by solving the gap equation
\begin{equation}
\frac{\partial \Omega_{\mu,B}^{\rm MFA,reg}}{\partial \bar{\sigma}} = 0 \ . \label{gapeq}
\end{equation}

Due to the presence of the magnetic field, the system becomes anisotropic, and the parallel and perpendicular pressures, $P_{\parallel}$ and $P_{\perp}$, are defined as

\begin{align}
P_{\parallel} &= -\Omega_{\mu,B}^{\rm MFA,reg},\\
P_{\perp} &= -\Omega_{\mu,B}^{\rm MFA,reg} - eB~\mathcal{M}, 
\end{align}
where $\mathcal{M}$ stand for the magnetization and is defined as
\begin{equation}
 \mathcal{M}= - \left(\frac{\partial~\Omega_{\mu,B}^{\rm MFA,reg}}{\partial (eB)} \right).    
\end{equation}

Following Refs.~\cite{Menezes:2015fla,Chaudhuri:2022oru,Goswami:2023eol}, we normalize the grand thermodynamic potential by subtracting, for each magnetic field strength, the regularized mean-field contribution evaluated at both zero temperature and zero chemical potential:
\begin{equation}
\Omega_{N} = \Omega_{\mu,B}^{\rm MFA,reg} - \Omega_{0,B}^{\rm MFA,reg},
\label{omegaN}
\end{equation}
where $\Omega_{N}$ denotes the normalized thermodynamic potential. 
This procedure ensures that the \textit{parallel pressure} vanishes in vacuum, providing a consistent reference point for different magnetic field values.

The corresponding normalized (or subtracted) parallel and perpendicular pressures are then defined as
\begin{equation}
P_{\parallel,N} = -\Omega_{N} = P_{\parallel}(\mu, B) - P_{\parallel}(0, B),
\end{equation}
and
\begin{equation}
P_{\perp,N} = P_{\parallel,N} - eB\, \mathcal{M}_{N},
\end{equation}
respectively. 
Here, $\mathcal{M}_{N}$ stands for the normalized magnetization, obtained from
\begin{equation}
\mathcal{M}_{N} = -\frac{\partial \Omega_{N}}{\partial (eB)}.
\label{magnet}
\end{equation}

We now turn to the analysis of the speed of sound in the presence of a magnetic field $ B $. 
In an anisotropic medium, it is necessary to distinguish between the longitudinal ($ C_{s,\parallel} $) and transverse ($ C_{s,\perp} $) components of the sound velocity, which are defined as \cite{Goswami:2023eol}
\begin{align}
C_{s,\parallel,N}^{2} &=  \frac{\partial P_{\parallel,N}}{\partial \varepsilon} \bigg|_{B}, 
\label{eq:c_parallel} \\
C_{s,\perp,N}^{2} &=  \frac{\partial P_{\perp,N}}{\partial \varepsilon} \bigg|_{B}, 
\label{eq:c_perp}
\end{align}
where $ \varepsilon $ denotes the energy density of the system. 
The latter can be expressed as
\begin{equation}
\varepsilon = \Omega_N + \mu \, n, 
\label{eq:energy_density}
\end{equation}
with $ n $ being the total quark number density (including both $u$ and $d$ quark contributions), obtained from the normalized thermodynamic potential through
\begin{equation}
n = -  \frac{\partial \Omega_N}{\partial \mu}\bigg|_{B}.
\label{eq:number_density}
\end{equation}

For completeness, we also introduce the total magnetic susceptibility, $ \chi_M $, which characterizes the response of the system to an external magnetic field:
\begin{equation}
\chi_M = -  \frac{\partial^2 \Omega_N}{\partial (eB)^2} \bigg|_{\mu}.
\label{eq:susceptibility}
\end{equation}

\section{Numerical Results}
\label{sect3}

In this section, we present the numerical results obtained within the nonlocal Nambu--Jona-Lasinio model in the presence of an external magnetic field. 
Our analysis focuses on the thermodynamic properties and anisotropic behavior of quark matter under different magnetic field strengths and chemical potentials. 
To perform the calculations, it is necessary to specify both the functional form of the nonlocal form factor $ g(p^2) $, introduced in Eq.~(\ref{funcg}), and the corresponding values of the model parameters. 
For simplicity we adopt a Gaussian ansatz,
\begin{equation}
g(p^2) = \exp\left(-\frac{p^2}{\Lambda^2}\right),
\label{eq:gaussian_formfactor}
\end{equation}
which allows the integral in Eq.~(\ref{funcg}) to be evaluated analytically and significantly reduces the computational cost of subsequent numerical calculations.

Within this framework, the constituent quark masses are determined as functions of the mean fields and the nonlocal form factor, as described in Ref.~\cite{GomezDumm:2017iex}
\begin{align}
M_{k,p_\parallel}^{\pm,f} = \ & (1-\delta_{k,-1})\, m_{c} \nonumber \\
&+
\bar{\sigma}\frac{\left( 1-|q_{f}B|/\Lambda^{2}\right)^{k_\pm}}{\left(1+|q_{f}B|/\Lambda^{2}\right)^{k_\pm+1}}
\,\exp(-p_\parallel^{2}/\Lambda^{2})\ .
\end{align}

We consider both the chiral limit and the case of finite $m_c$, fixing the model parameters 
to reproduce physical observables at $\mu = B = 0$ and adopting $-\langle \bar{q}q\rangle^{1/3} = 230~\mathrm{MeV}$ in both cases. 
In the chiral limit, we obtain 
$\Lambda = 684~\mathrm{MeV}$ and $G\Lambda^2 = 23.42$, for $f_{\pi,\mathrm{ch}} = 90~\mathrm{MeV}$~\cite{Ferraris:2021vzm}, 
while for finite $m_c$ we obtain 
$m_c = 6.5~\mathrm{MeV}$, $\Lambda = 678~\mathrm{MeV}$, and $G\Lambda^2 = 23.66$~\cite{GomezDumm:2017iex}, 
for $m_\pi = 139~\mathrm{MeV}$ and $f_\pi = 92.4~\mathrm{MeV}$.

The gap equation, Eq.~(\ref{gapeq}), is then solved numerically for each combination of chemical potential $\mu$ and magnetic field $B$. 
In regions where multiple solutions exist, the physically realized solution is identified as the one that minimizes the thermodynamic potential, corresponding to the absolute ground state of the system. 
This procedure ensures thermodynamic consistency and selects the stable configuration of the quark condensates for the given external conditions.

\subsection{Chiral limit case}

In our previous study~\cite{Ferraris:2025bzb}, we performed a preliminary analysis of pressure anisotropy in cold quark matter under strong magnetic fields in the chiral limit, using an alternative parameter set and a thermodynamic potential normalized by subtracting only the vacuum contribution at $B=0$. 
In the present work, still within the chiral limit, we extend that analysis to include additional thermodynamic and chiral properties. 
Here we adopt a new parameter set and employ the normalization scheme introduced in Eq.~(\ref{omegaN}), while keeping the same chiral condensate as in the finite-mass case to be discussed in Sect.~\ref{finite_mass}. 
This setup provides a consistent baseline for comparison and allows a unified assessment of the effects of strong magnetic fields on the anisotropic pressures as well as on other relevant thermodynamic quantities.

\begin{widetext}

\begin{figure}[H]
\centering 
\includegraphics[width=0.78\textwidth]{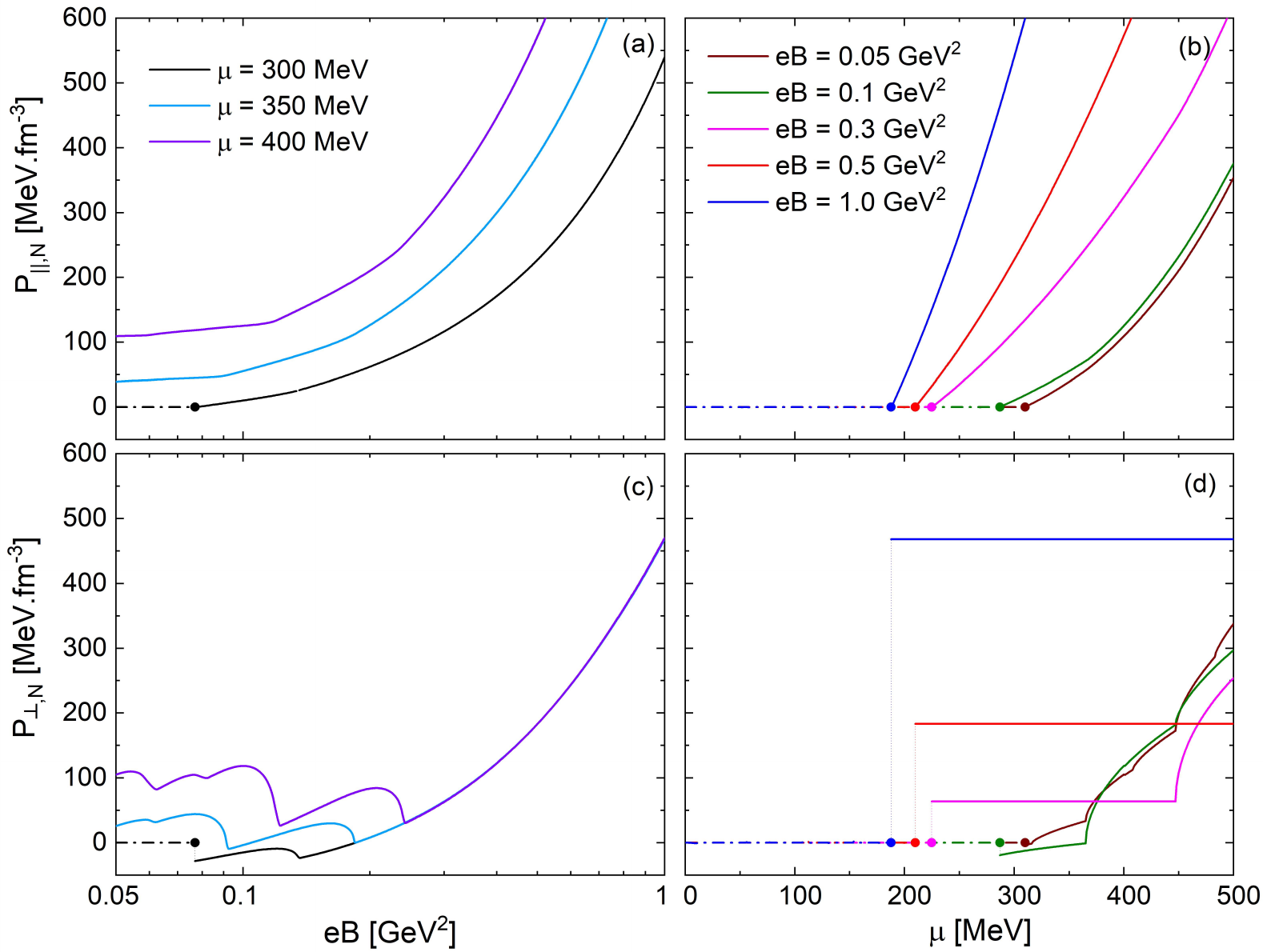}
\caption{Normalized longitudinal ($P_{\parallel}$) and transverse ($P_{\perp}$) pressures as functions of the magnetic field strength $eB$ and the quark chemical potential $\mu$ in the chiral limit.  In all cases, dashed curves correspond to the $\chi SB$ phase and solid curves to the $\chi SR$ phase and the first-order chiral phase transition is marked by a dot.}  
\label{pressure_perp_mu_chi}
\end{figure}   

\end{widetext}

We begin our analysis by presenting in Fig.~\ref{pressure_perp_mu_chi} the normalized longitudinal $P_{\parallel,N}$ and transverse $P_{\perp,N}$ pressures as functions of the magnetic field strength $eB$ and the quark chemical potential $\mu$ (hereafter the subscript $N$ will be omitted). Dashed lines correspond to the broken phase of the chiral symmetry ($\chi SB$), while the solid lines represent the restored phase ($\chi SR$). 

The values of $\mu$ shown in this figure were selected around the critical chemical potential, $\mu_c \sim 320$~MeV, obtained at vanishing magnetic field in Ref.~\cite{Ferraris:2021vun}, in order to examine the system’s behavior in the vicinity of the phase transition.

Fig.~\ref{pressure_perp_mu_chi} (a) shows the longitudinal pressure $P_{\parallel}$ as a function of $eB$ for three representative chemical potentials. For the lowest $\mu$, the system starts in the $\chi SB$ phase and undergoes a first-order phase transition to the $\chi SR$ phase, indicated by the dot on the curve. 
At this transition, $P_{\parallel}$ exhibits a change from vanishing values in the $\chi SB$ phase (since the normalization is taken with respect to the thermodynamic pressure) to a rapidly increasing behavior in the $\chi SR$ phase. 
For higher values of $\mu$, the system lies entirely in the phase $\chi SR$, where the pressure grows monotonically with $eB$. This stronger growth reflects both the reduced effective quark mass and the larger filling of Landau levels, which enhance the kinetic contribution along the direction of the field.

In Fig.~\ref{pressure_perp_mu_chi} (b), $P_{\parallel}$ is plotted as a function of $\mu$ for several fixed magnetic fields. At low chemical potential, the pressure is zero in the $\chi SB$ phase, while at larger $\mu$ it increases monotonically in the $\chi SR$ phase. The transition between phases is clearly manifested as a steep increase in pressure, which becomes more pronounced as $eB$ increases. This behavior shows that the magnetic field enhances the longitudinal pressure once the system enters the restored phase, while the critical chemical potential for chiral symmetry restoration decreases with increasing $eB$, signaling the occurrence of $\mu$IMC as reported in \cite{Ferraris:2021vun}.

Fig.~\ref{pressure_perp_mu_chi} (c) and (d) display the behavior of transverse pressure $P_{\perp}$. In panel (c), $P_{\perp}$ is plotted as a function of $eB$ for the same values of $\mu$ considered in (a). For low magnetic fields, $P_{\perp}$ approaches the corresponding longitudinal pressure $P_{\parallel}$. For $\mu = 300$ MeV, a region in low fields corresponds to the $\chi SB$ phase (dashed), followed by a first-order transition to the $\chi SR$ phase. In contrast, for the other values of the chemical potential, the system remains completely in the $\chi SR$ phase. Contrary to the longitudinal case, here the pressure initially decreases as the magnetic field increases, due to the progressive suppression of transverse-quark motion by Landau quantization. At sufficiently strong fields ($eB \gtrsim 0.2$ GeV$^2$), $P_{\perp}$ begins to grow, and the curves for different $\mu$ essentially overlap. At intermediate fields, $P_{\perp}$ can become negative, indicating the dominance of the magnetization term over the thermodynamic contribution. This effect is particularly pronounced in the $\chi SR$ phase. Oscillations are also visible, arising from the discrete filling of Landau levels, which directly affect the magnetization. Comparison between phases shows that, at the same $eB$, the $\chi SR$ phase maintains higher transverse pressures than the broken phase.

Finally, Fig.~\ref{pressure_perp_mu_chi} (d) presents the perpendicular pressure $P_{\perp}$ as a function of the chemical potential $\mu$ for different fixed values of the magnetic field $eB$. 
The transition between the two phases is marked by a clear change in both the slope and magnitude of $P_{\perp}$, with more pronounced discontinuities observed at higher magnetic fields.
For weaker magnetic fields, once the system enters the $\chi SR$ phase, the curves of $P_{\perp}(\mu)$ exhibit a gradual, step-like behavior, reflecting the sequential filling of the Landau levels. A key observation is that for the three strongest magnetic fields considered ($eB = 0.3, 0.5, 1.0~\text{GeV}^2$), the magnetization remains constant and non-zero in the restored phase. The numerical results show that magnetization reaches a field-dependent plateau in the restored phase for $eB \geq 0.3~\text{GeV}^2$. This behavior can be understood from the dimensional reduction of quark motion in a strong magnetic field. For sufficiently high $eB$, the quarks become confined to the LLL, their dynamics becoming effectively one-dimensional along the direction of the magnetic field. In this $(1+1)$-dimensional regime, the energy spectrum of the quarks and consequently the system's thermodynamic properties, such as magnetization, become less sensitive to changes in the chemical potential $\mu$, leading to the observed plateau.

It is instructive to compare our results with those shown in Fig.~11(a) of Ref.~\cite{Chaudhuri:2022oru}. 
The behavior of the longitudinal pressure $P_{\parallel}$ as a function of $eB$ is qualitatively 
similar in both cases, increasing monotonically with the magnetic field. For the transverse 
pressure $P_{\perp}$, we also find very good agreement: in both calculations $P_{\perp}$ 
initially decreases with $eB$ due to Landau quantization, and for sufficiently strong fields it 
starts to rise again. Moreover, the oscillatory pattern associated with the discrete filling of 
Landau levels is clearly visible in our results, in line with the features reported in 
Ref.~\cite{Chaudhuri:2022oru}. Thus, despite differences in the model setup and regularization schemes, 
our findings are consistent with the general trends obtained in Ref.~\cite{Chaudhuri:2022oru}.

\begin{widetext}

\begin{figure}[H]
\centering 
\includegraphics[width=0.78\textwidth]{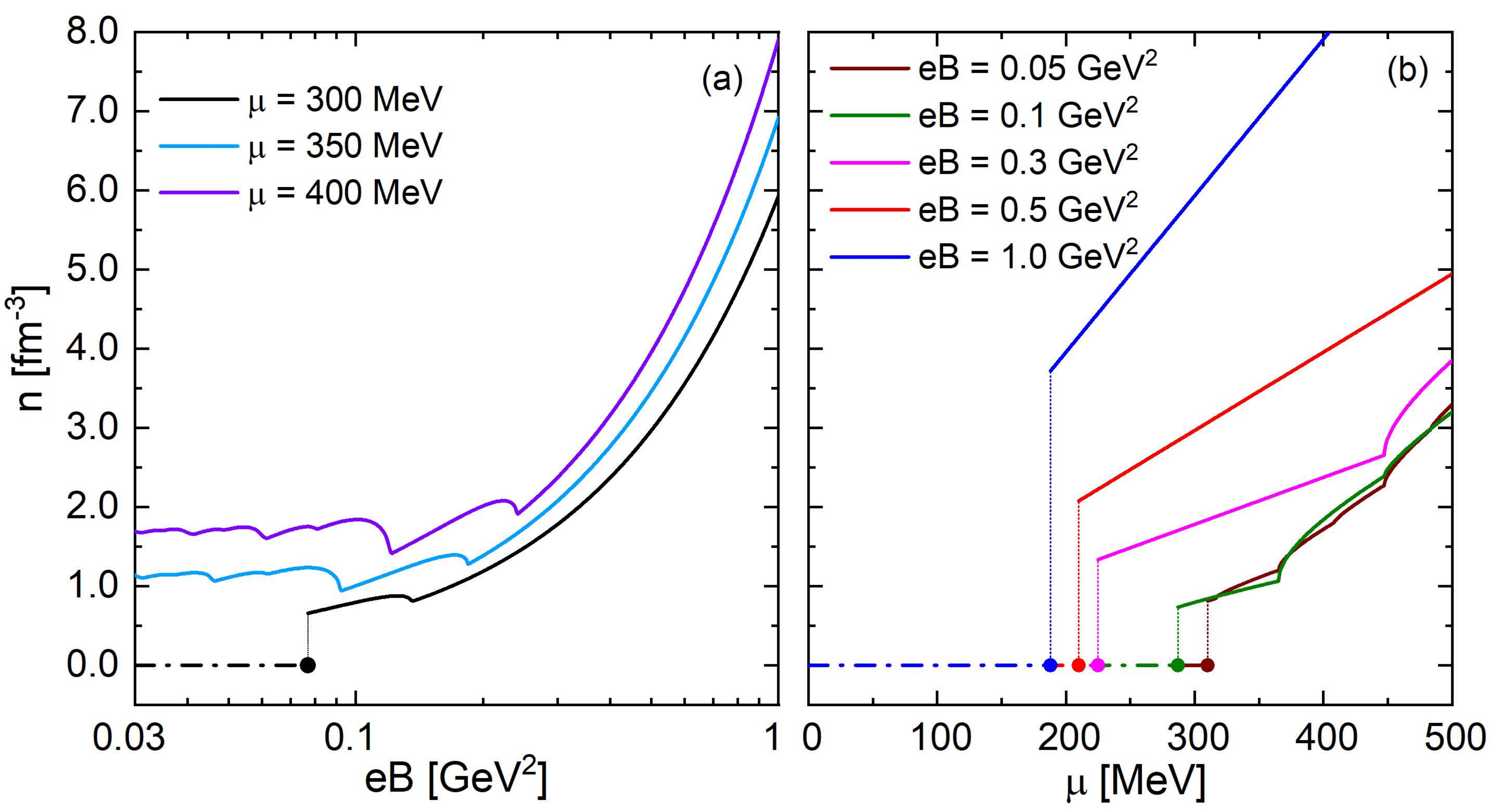}
\caption{Quark number density $n$ as a function of (a) the magnetic field $eB$ and (b) the quark chemical potential $\mu$. Results are shown for three representative chemical potentials, as indicated in (a) and several magnetic field strengths in (b).}
\label{fig:density_merge_chiral}
\end{figure}   

\end{widetext}

Figure~\ref{fig:density_merge_chiral}(a) shows the quark number density $n$ as a function of the magnetic field $eB$ for the same three representative chemical potentials as before. 
The density increases with $eB$ due to the enhanced Landau-level degeneracy, and the effect becomes stronger at larger $\mu$, where the system lies in the $\chi SR$ phase and additional longitudinal modes are accessible. 
At low and intermediate magnetic fields, the discrete population of Landau levels produces visible modulations in $n$, which gradually fade as the system enters the LLL regime, where $n$ grows monotonically with $eB$.

Figure~\ref{fig:density_merge_chiral}(b) displays $n$ as a function of the chemical potential $\mu$ for fixed magnetic fields. 
In the chiral limit, the order parameter of the system is the chiral condensate, which vanishes in the $\chi SR$ phase. 
Since it remains constant (and zero) throughout this phase, the identification of the different regions associated with the successive filling of Landau levels must rely on the behavior of the density. 
The corresponding transitions manifest as variations of $n(\mu)$—continuous changes in slope rather than discontinuous jumps—indicating that they are of second order. 
These features, characteristic of the quantized structure of the Landau levels, become less pronounced as $eB$ increases, when the system approaches the LLL-dominated regime.

\begin{widetext}

\begin{figure}[H]
\centering 
\includegraphics[width=0.78\textwidth]{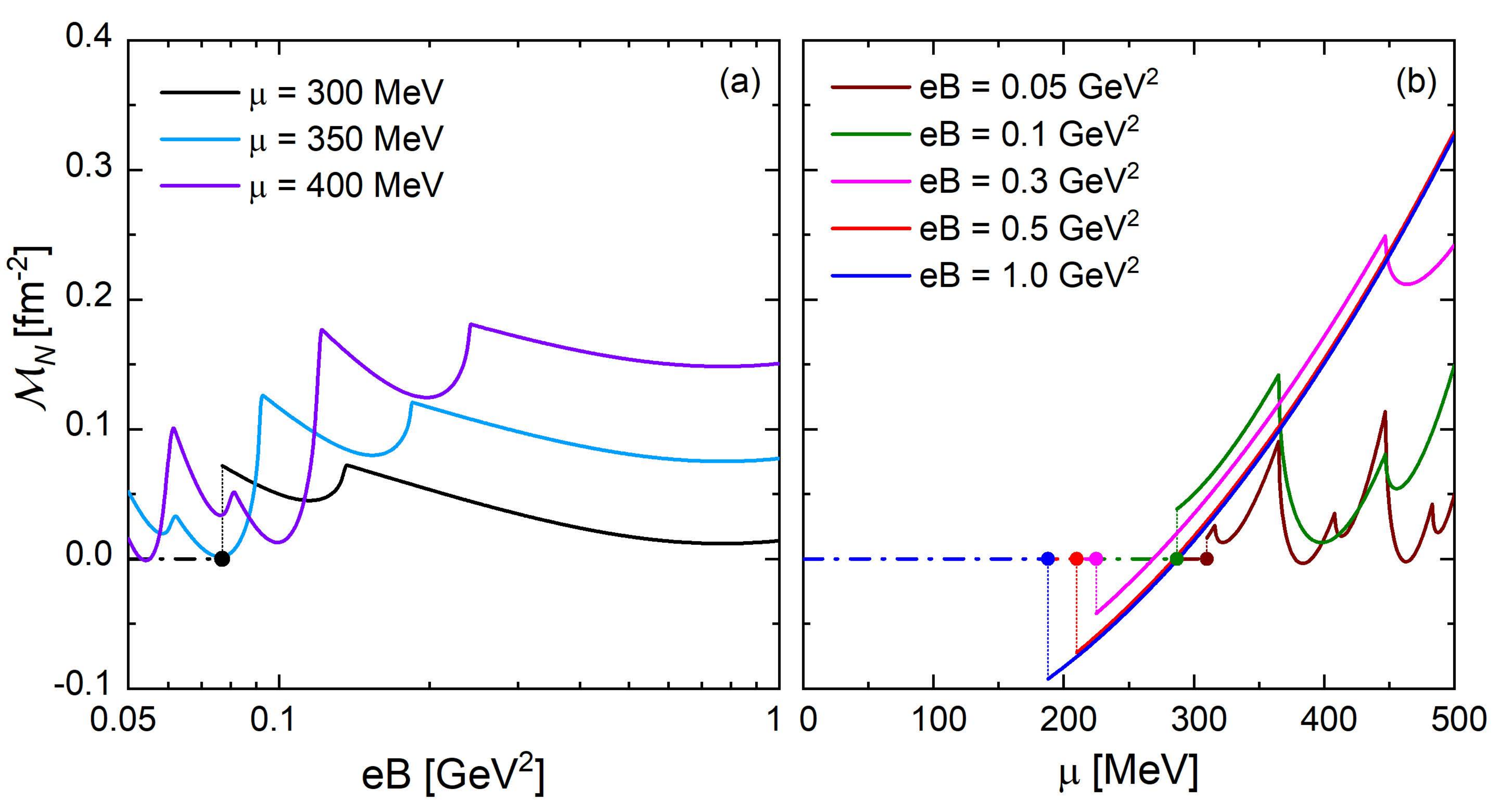}
\caption{Normalized magnetization as function of (a) the magnetic field $eB$ and (b) the quark chemical potential $\mu$, for the chiral limit.}
\label{fig:magnetiz_merge_chiral}
\end{figure}   

\end{widetext}

Figure~\ref{fig:magnetiz_merge_chiral}(a) presents the normalized magnetization 
$\mathcal{M}_N = -\partial\Omega_N / \partial(eB)$ as a function of $eB$ for three fixed chemical potentials. 
For the lowest $\mu$, the system remains in the $\chi SB$ phase at weak fields, where $\mathcal{M}_N \simeq 0$. 
As $eB$ increases, a first-order chiral transition occurs and $\mathcal{M}_N$ becomes finite, exhibiting oscillations due to the sequential filling of Landau levels. 
These de Haas--van Alphen--like oscillations are gradually damped as $eB$ increases, and $\mathcal{M}_N$ approaches a nearly constant value in the LLL-dominated regime.
For the lowest fields, our results are consistent with those of Fig.~11(b) of Ref.~\cite{Chaudhuri:2022oru}, which explored a range up to $eB \simeq 0.16~\mathrm{GeV}^2$.  

Figure~\ref{fig:magnetiz_merge_chiral}(b) shows $\mathcal{M}_N$ as a function of $\mu$ for representative magnetic field strengths. 
The magnetization vanishes in the $\chi SB$ phase and becomes nonzero after the first-order transition (indicated by dots). 
Within the $\chi SR$ phase, $\mathcal{M}_N$ grows with $\mu$, reaches a maximum near each Landau-level threshold, and then decreases as the next level begins to populate. 

\begin{figure}[H]
\centering 
\includegraphics[width=0.48\textwidth]{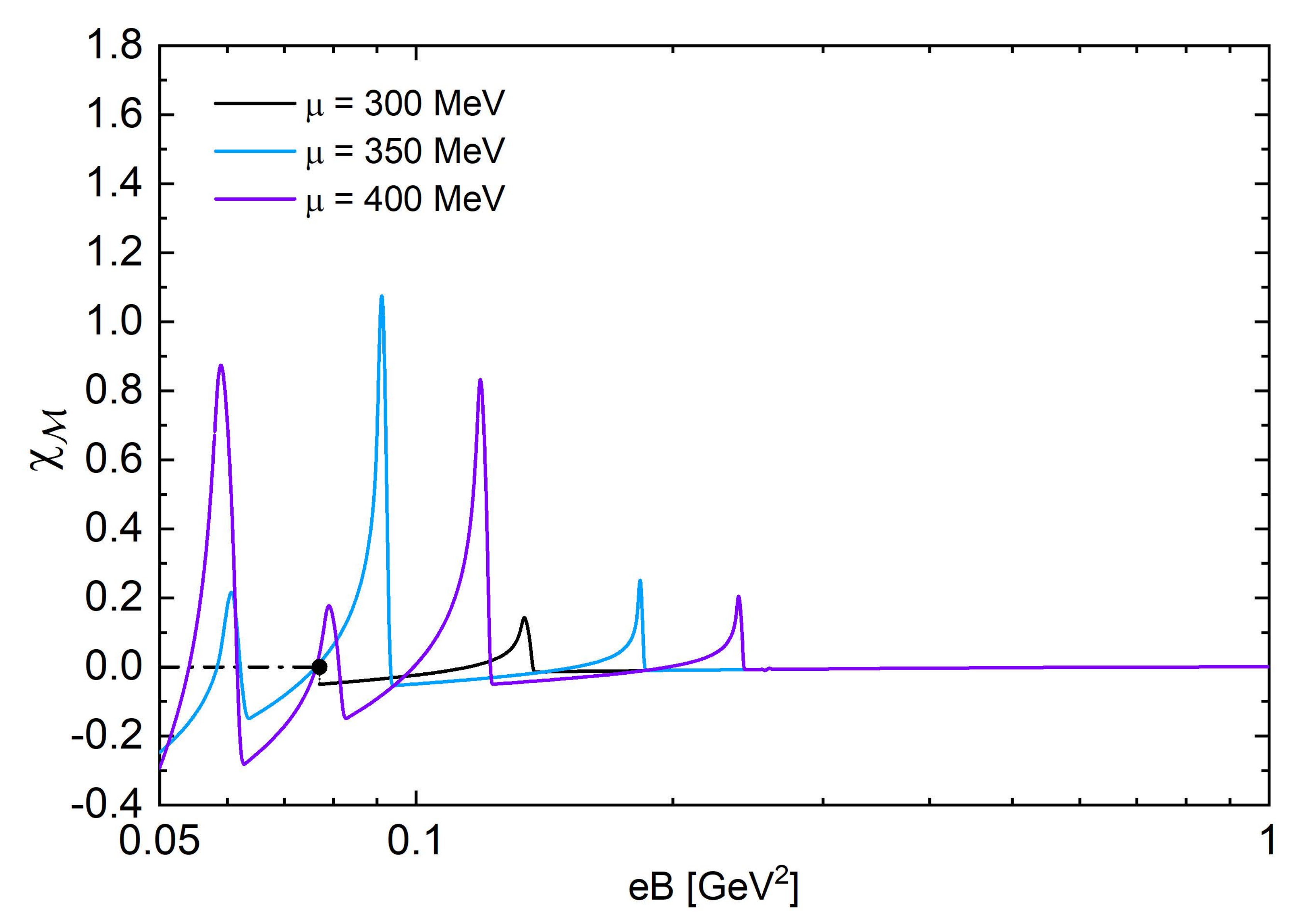}
\caption{Magnetic susceptibility as function of magnetic field $eB$ for different values of quark chemical potential $\mu$, within the chiral limit.}
\label{fig:suscept_chiral}
\end{figure}

Next, we present the normalized magnetic susceptibility in Fig.~\ref{fig:suscept_chiral} for the same fixed values of the chemical potential considered before. At low magnetic fields, the susceptibility exhibits pronounced oscillations that alternate between positive and negative values. These oscillations originate from the Landau-level structure: whenever a new Landau level becomes accessible, the density of states increases abruptly, leading to an enhancement of the response, while in between thresholds the signal decreases and may even turn negative. As a result, the system displays a characteristic oscillatory pattern whose amplitude gradually diminishes with increasing $\mu$, since a denser medium reduces the relative impact of individual Landau-level thresholds on the thermodynamic quantities. For stronger magnetic fields, the spacing between Landau levels grows and eventually only the lowest Landau level remains populated. In this LLL-dominated regime, the oscillations fade away and the curves tend to stabilize. While the plotted range shows the susceptibility approaching zero, a positive trend emerges at high $eB$, consistent with the behavior expected when the system becomes dominated by the lowest Landau level. Thus, the behavior evolves from strong oscillations at low $eB$ to a regime where the susceptibility becomes small and positive at high fields. Our results are in qualitative agreement with previous analyses. In particular, the oscillatory behavior at low $eB$ and its suppression at higher fields closely resemble the pattern reported in Fig.~11(c) of Ref.~\cite{Chaudhuri:2022oru}, supporting the consistency of our findings with earlier effective-model studies.

Now we turn to the analysis of the equation of state and the speed of sound in the $\chi SR$ phase, focusing on the regime in which the system presents a Landau-level structure. In this region, the system undergoes successive phases corresponding to the filling of different Landau levels, separated by de Haas–van Alphen–like transitions. These transitions induce abrupt changes in the slopes of thermodynamic quantities, giving rise to the oscillatory behavior observed in both the pressures and the speed of sound.

\begin{widetext}

\begin{figure}[H]
\centering 
\includegraphics[width=0.78\textwidth]{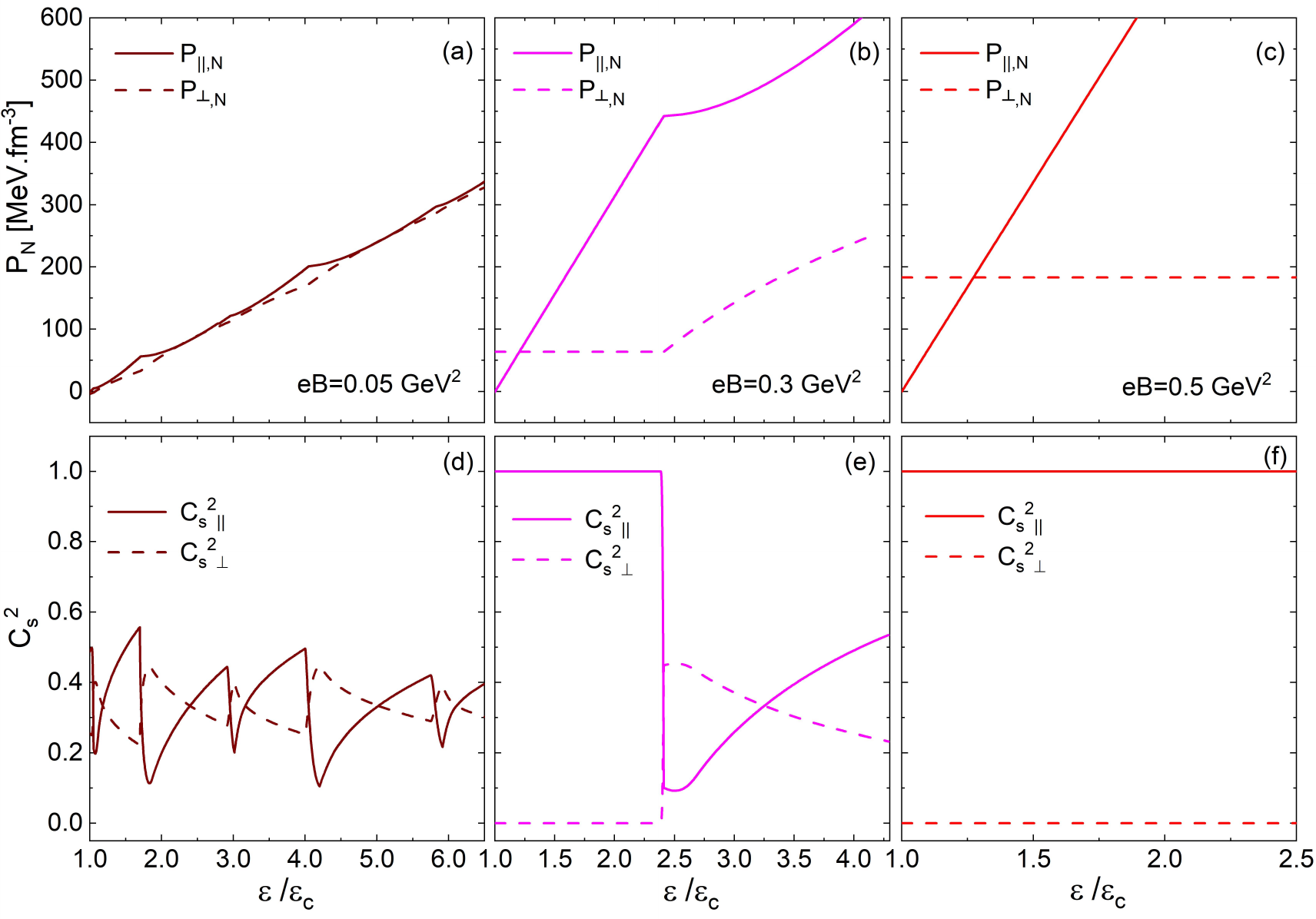}
\caption{Normalized longitudinal (solid) and transverse (dashed) pressures (a)-(c) and corresponding squared sound velocities $c_s^2$ (d)-(f) as functions of the dimensionless energy density $\varepsilon/\varepsilon_c$. Results are shown for three magnetic field strengths: $eB=0.05$ GeV$^2$ ((d), $\varepsilon_c=184.1$ MeV/fm$^3$), $eB=0.3$ GeV$^2$ ((e), $\varepsilon_c=313.7$ MeV/fm$^3$), and $eB=0.5$ GeV$^2$ ((f), $\varepsilon_c=672.0$ MeV/fm$^3$). }
\label{Fig:EoS_Cs2_chiral}
\end{figure}   

\end{widetext}

Figure~\ref{Fig:EoS_Cs2_chiral}(a)–(c) shows normalized longitudinal and transverse pressures, while panels (d)–(f) show the corresponding squared sound velocities, as functions of the dimensionless energy density $\varepsilon/\varepsilon_c$, where $\varepsilon_c$ denotes the energy density in chiral phase transition for each magnetic field. Solid lines correspond to longitudinal quantities, dashed lines to transverse ones, and results are shown for $eB = 0.05$, $0.3$, and $0.5~\mathrm{GeV^2}$.

The longitudinal pressure $P_{\parallel}$ increases monotonically with $\varepsilon$, with a slope that steepens as $eB$ grows, indicating a stiffening of the equation of state. Kink-like features appear whenever a new Landau level is populated, reflecting abrupt changes in the density of states. In the strong-field limit, where only the lowest Landau level contributes, transverse motion is frozen, and the system effectively behaves as $(1+1)$-dimensional. In this regime, $P_{\parallel} \simeq \varepsilon$ and $C_{s,\parallel}^2 \to 1$, consistent with a free gas of massless fermions constrained to move along the field.

Peaks in $C_{s,\parallel}^2$ at intermediate densities arise from the threshold population of higher Landau levels: each newly accessible level enhances $dP_{\parallel}/d\varepsilon$, producing a peak. At low fields, many levels contribute, leading to numerous but smaller peaks; at stronger fields, fewer levels produce sharper, more widely spaced peaks, yet the density range remains similar, reflecting the universal pattern of Landau quantization.

\begin{figure}[H]
\centering 
\includegraphics[width=0.43\textwidth]{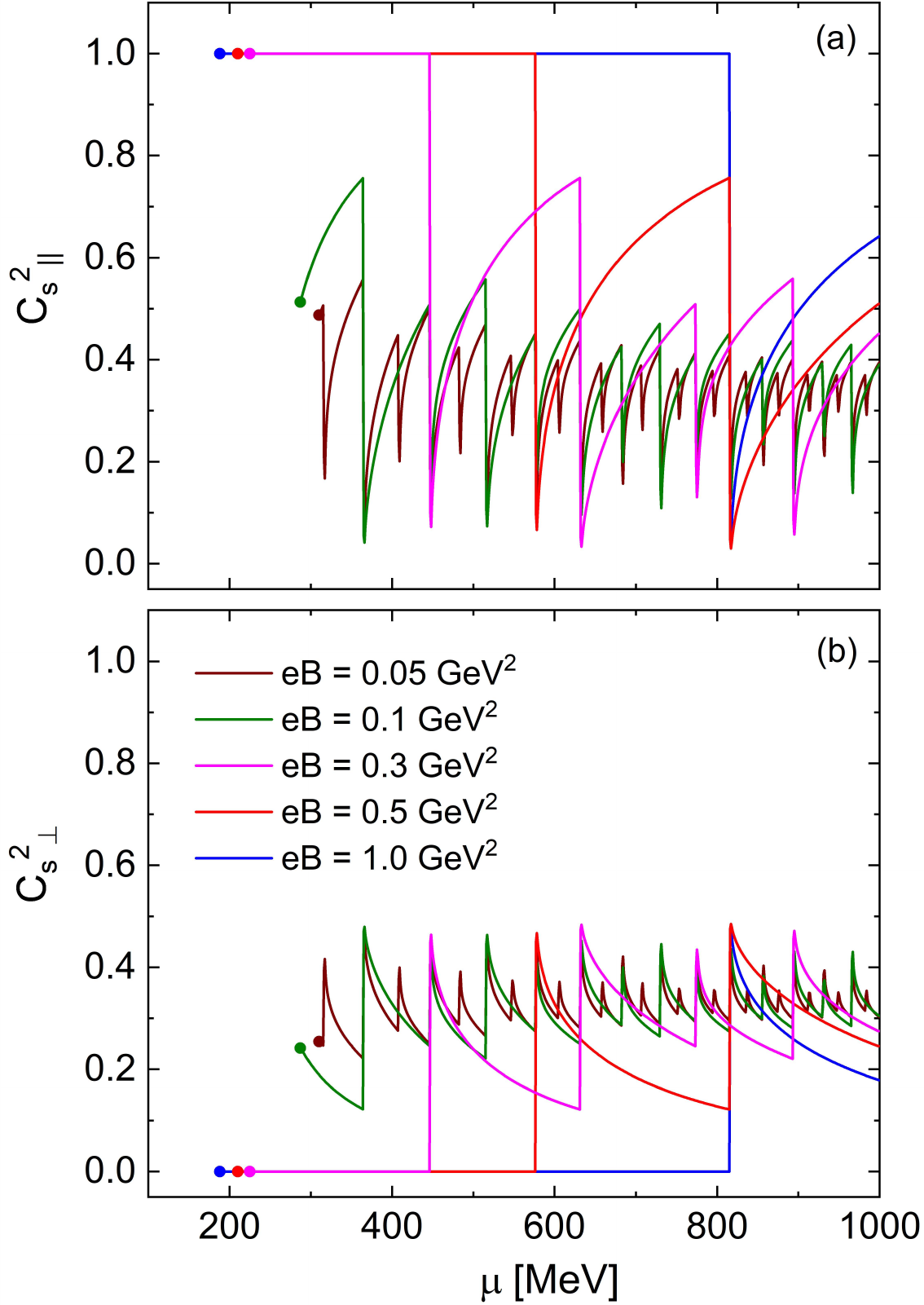}
\caption{Square of the longitudinal (a) and perpendicular (b) 
sound velocities, $C_s^2$, as a function of the chemical potential $\mu$ for different values of the external magnetic field $eB$.}
\label{Fig:Cs2_vs_mu_chiral}
\end{figure} 

In the LLL regime, transverse dynamics are nearly frozen, maximizing the anisotropy: $P_{\parallel}$ continues to increase, while $P_{\perp}$ saturates. The corresponding transverse sound velocity $C_{s,\perp}^2$ decreases with $eB$ and vanishes at strong fields. Thus, the system behaves effectively as one-dimensional, with excitations propagating along the magnetic field ($C_{s,\parallel}^2 \to 1$) and transverse propagation suppressed ($C_{s,\perp}^2 \to 0$).

To further extend the analysis in the chiral limit, Figure~\ref{Fig:Cs2_vs_mu_chiral} (a) and (b) show the squared speed of sound along ($C_{s,\parallel}^2$) and perpendicular ($C_{s,\perp}^2$) to the magnetic field, respectively, as functions of the quark chemical potential $\mu$ for several field strengths, still within the $\chi SR$ phase.

In the strong-field regime dominated by the LLL, quark motion is effectively $(1+1)$-dimensional, resulting in maximal anisotropy: $C_{s,\parallel}^2$ approaches unity, saturating the causal limit, while $C_{s,\perp}^2$ vanishes due to the suppression of transverse dynamics. Beyond this regime, both velocities oscillate around the conformal limit $1/3$, reflecting de Haas–van Alphen–like transitions associated with the successive filling of Landau levels. The different electric charges of $u$ and $d$ quarks produce distinct Landau-level spacings, generating complex oscillation patterns: each newly populated level abruptly increases the density of states, producing peaks in $C_{s,\parallel}^2$ and corresponding dips in $C_{s,\perp}^2$, particularly pronounced at intermediate $\mu$. The spacing between consecutive peaks scales with $eB$, consistent with the increasing number of de Haas–van Alphen–like transitions at lower fields.  

\subsection{Finite current quark mass case}
\label{finite_mass}

\begin{widetext}

\begin{figure}[H]
\centering 
\includegraphics[width=0.78\textwidth]{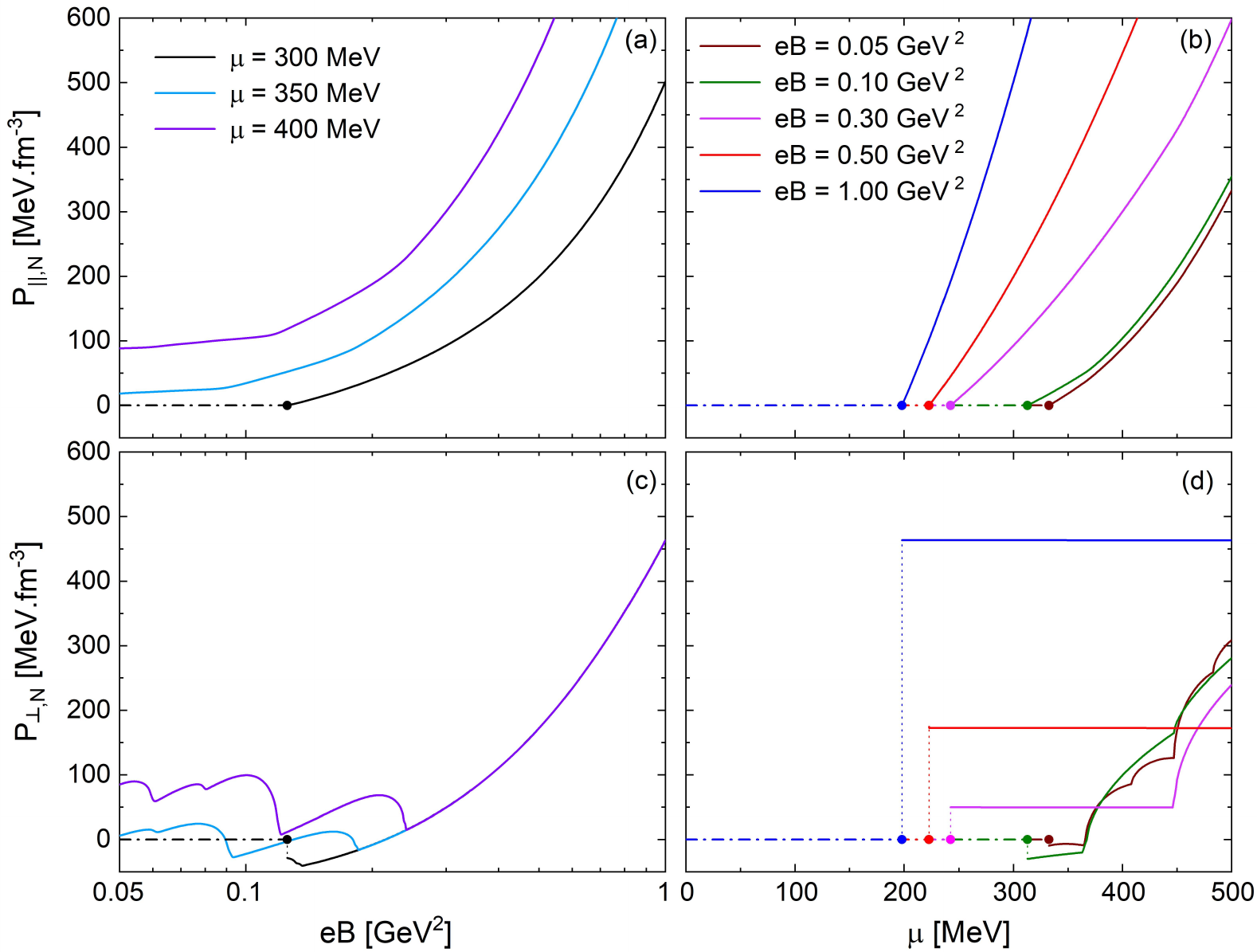}
\caption{ Normalized longitudinal ($P_{\parallel}$) and transverse ($P_{\perp}$) pressures as functions of the magnetic field $eB$ and quark chemical potential $\mu$ for finite current quark masses. Dashed lines correspond to the broken phase of the chiral symmetry, and solid lines correspond to the restored phase of the chiral symmetry. (a) $P_{\parallel}$ vs $eB$ and (c) $P_{\perp}$ vs $eB$ are shown for three fixed values of $\mu$, while (b) $P_{\parallel}$ vs $\mu$ and (d) $P_{\perp}$ vs $\mu$ are shown for several fixed values of $eB$, as in previous figures.}
\label{presiones_non_chiral}
\end{figure}   

\end{widetext}

Having established the baseline behavior in the chiral limit, we now extend the analysis to finite current quark masses. The results are presented in Fig.~\ref{presiones_non_chiral}(a)–(b) and (c)–(d) in the same manner as for the chiral case, showing the normalized longitudinal and transverse pressures as functions of the magnetic field strength $eB$ and the quark chemical potential $\mu$, respectively. Dashed lines indicate the chirally broken phase, while solid lines correspond to the regime of partial chiral symmetry restoration; however, for brevity, we will continue to denote it as $\chi SR$ throughout the text.

As in the chiral limit case, the chemical potentials in  Fig.~\ref{presiones_non_chiral}(a) and (c) were chosen in the vicinity of the critical value at low magnetic fields ($\mu_c \sim 340$~MeV~\cite{Ferraris:2021vun}). 

Panels (a) and (b) illustrate $P_{\parallel}$ as a function of $eB$ and $\mu$, respectively. The behavior is qualitatively the same as in the chiral limit case: $P_{\parallel}$ grows with both the magnetic field and chemical potential, reflecting the progressive filling of Landau levels and the reduction of the effective quark mass. The longitudinal pressure is enhanced by stronger fields, and the transition between phases shows the same steep increase as in the chiral case.

Panels (c) and (d) show the transverse pressure $P_{\perp}$. Like in the chiral limit, $P_{\perp}$ is generally suppressed relative to $P_{\parallel}$ due to Landau quantization. Oscillations from discrete Landau levels are still visible, and strong magnetic fields can still drive $P_{\perp}$ negative, though less prominently than for massless quarks.

Overall, Fig.~\ref{presiones_non_chiral} demonstrates that the magnetic field-induced anisotropy behavior persists with finite quark masses: the longitudinal pressure is enhanced while the transverse pressure is reduced, and, the qualitative behavior closely follows that of the chiral limit.

\begin{widetext}

\begin{figure}[H]
\centering 
\includegraphics[width=0.78\textwidth]{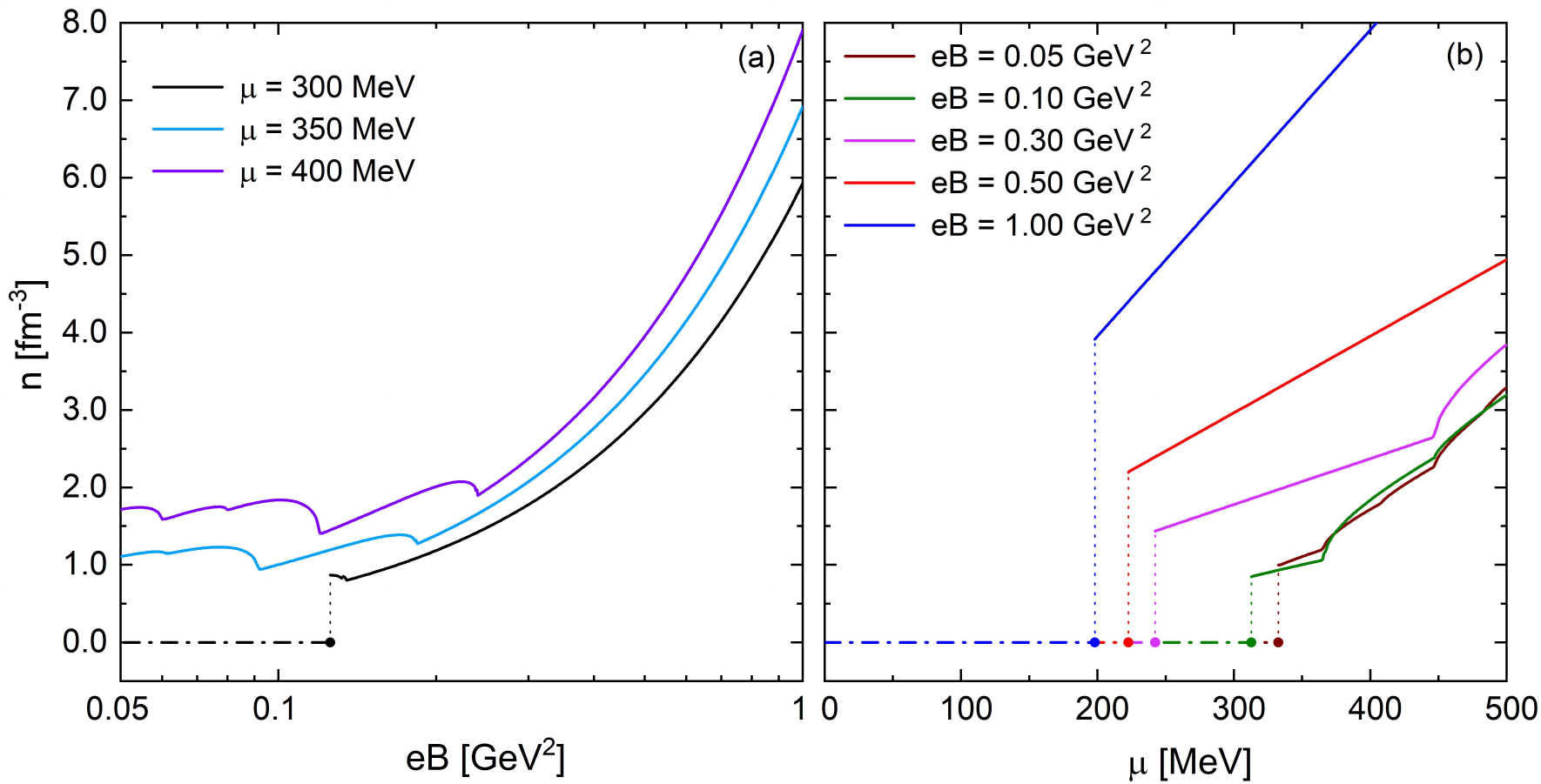}
\caption{Quark number density $n$ as function of magnetic field $eB$ (a) and quark chemical potential $\mu$ (b).}
\label{fig:density_merge}
\end{figure}   

\end{widetext}

We next examine the quark number density $n$, a quantity of central importance for determining the equation of state and the related thermodynamic observables. This analysis extends our previous study in the chiral limit~\cite{Ferraris:2025bzb} to the case of finite current quark masses, allowing for a direct comparison between both regimes. The corresponding results are displayed in Fig.~\ref{fig:density_merge}. 

Figure~\ref{fig:density_merge}(a) shows the quark number density $n$ as a function of $eB$ for the same chemical potentials used in the chiral-limit analysis. 
In all cases, $n$ increases with $eB$ because of the growing Landau-level degeneracy, and the effect is more pronounced at larger $\mu$, where the system lies in the $\chi SR$ phase. 
For sufficiently strong fields, only the LLL contributes, leading to a monotonic increase of $n$ with $eB$. 
At lower fields, the discrete filling of higher Landau levels produces oscillatory patterns that leave clear imprints on the thermodynamic quantities.

Figure~\ref{fig:density_merge}(b) presents $n$ as a function of $\mu$ for several fixed magnetic fields. 
At low $eB$, kink-like features appear as new Landau levels begin to populate, reflecting the discrete nature of transverse quantization. 
As $eB$ increases, the separation between levels increases, signaling the approach to the LLL regime.

Compared with the chiral-limit case, the qualitative trends remain the same: magnetic fields enhance the quark density, and this effect becomes more pronounced at higher chemical potentials. 
However, the presence of a finite current-quark mass alters the nature of the Landau-level transitions. 
Since the chiral condensate no longer vanishes in the $\chi SR$ phase, its discontinuous behavior across successive levels renders these transitions weakly of the first order. 
As a consequence, small jumps appear in the quark number density and related thermodynamic quantities, reflecting the discrete restructuring of the Fermi surface at each Landau-level threshold. 
The monotonic increase of $n$ with $eB$ also contributes to the overall growth of the energy density and plays a significant role in shaping the equation of state and the magnetization at strong fields.

Building on the analysis of the quark number density, which provides essential input for the equation of state, we now turn to the magnetization, a quantity that also plays a central role in determining the system’s thermodynamic response to a magnetic field. Although $\mathcal{M}$ has already been implicitly involved in the evaluation of the transverse pressure and the equation of state, we now examine its behavior explicitly. We focus on the normalized magnetization as a function of both the chemical potential $\mu$ and the magnetic field $eB$. Magnetization encodes the response of the system to the external field and is crucial for understanding the magnetic properties of quark matter, contributing directly to the transverse pressure and thus influencing the anisotropy of the equation of state. 
\begin{widetext}

\begin{figure}[H]
\centering 
\includegraphics[width=0.78\textwidth]{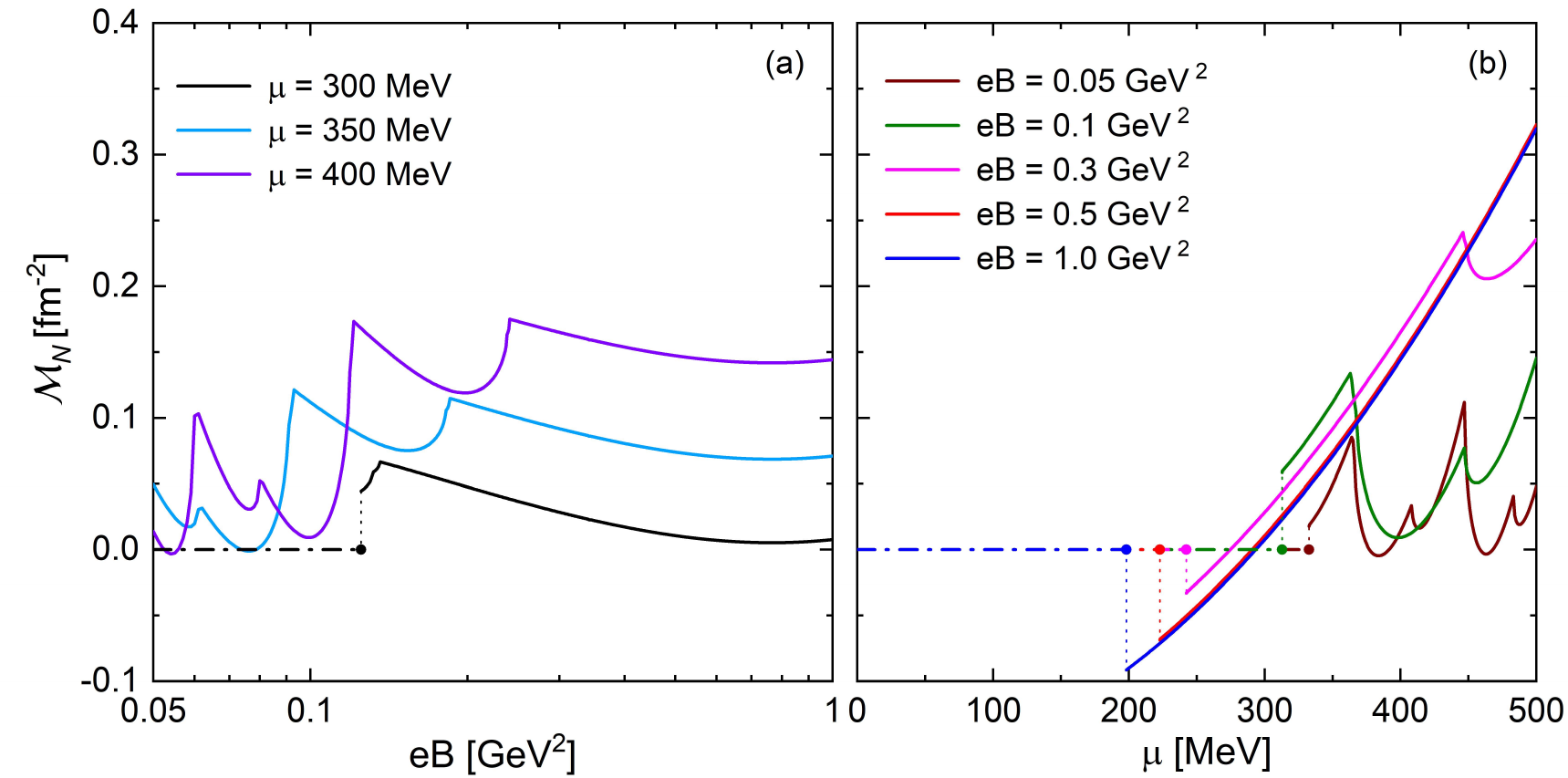}
\caption{Normalized magnetization as function of (a) the magnetic field $eB$ and (b) the quark chemical potential $\mu$.}
\label{fig:magnetiz_merge}
\end{figure}   
\end{widetext}

\begin{figure}[H]
\centering 
\includegraphics[width=0.48\textwidth]{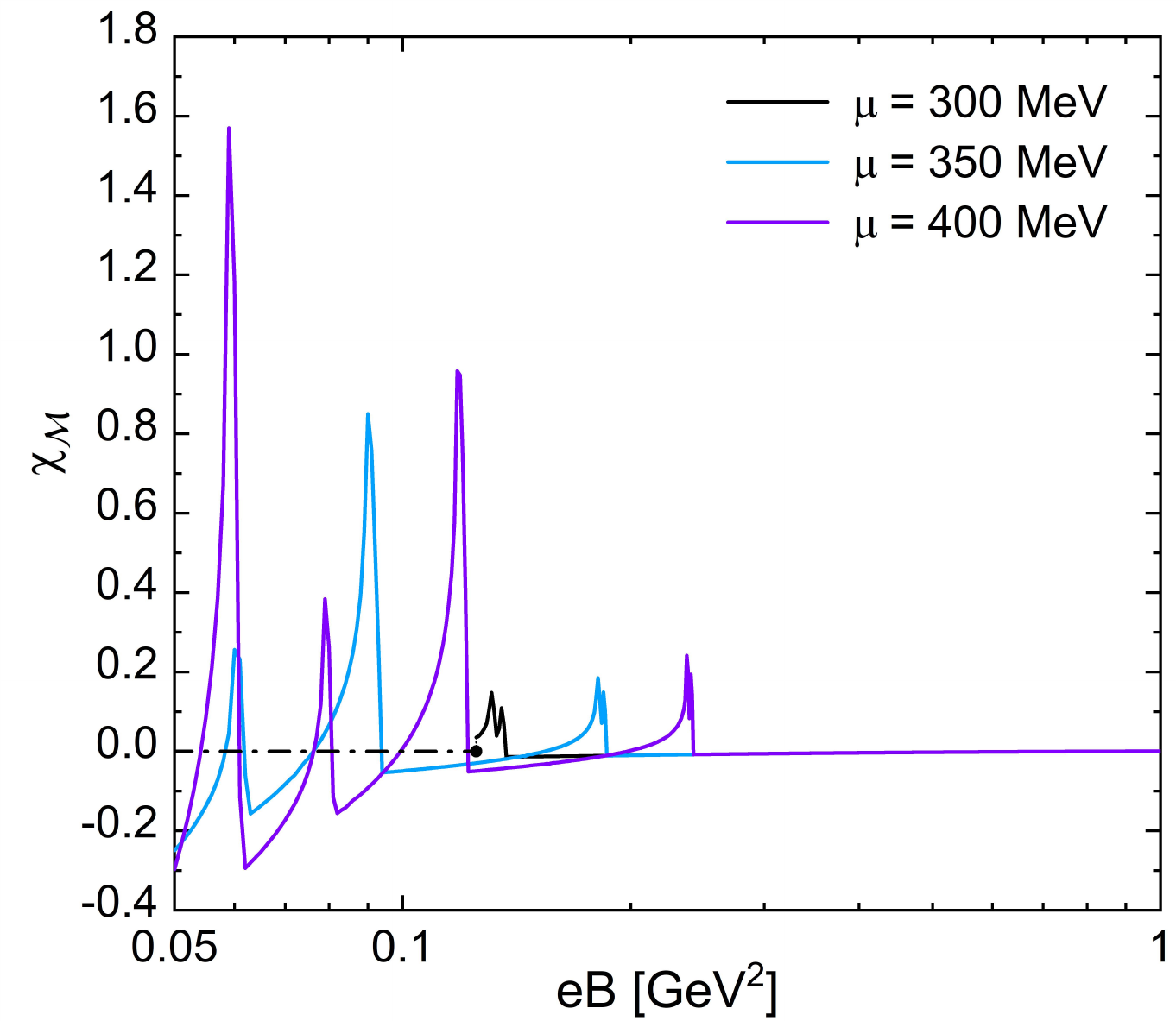}
\caption{Magnetic susceptibility as function of magnetic field $eB$ for different values of quark chemical potential $\mu$.}
\label{fig:suscept}
\end{figure}

\begin{widetext}

\begin{figure}[H]
\centering 
\includegraphics[width=0.78\textwidth]{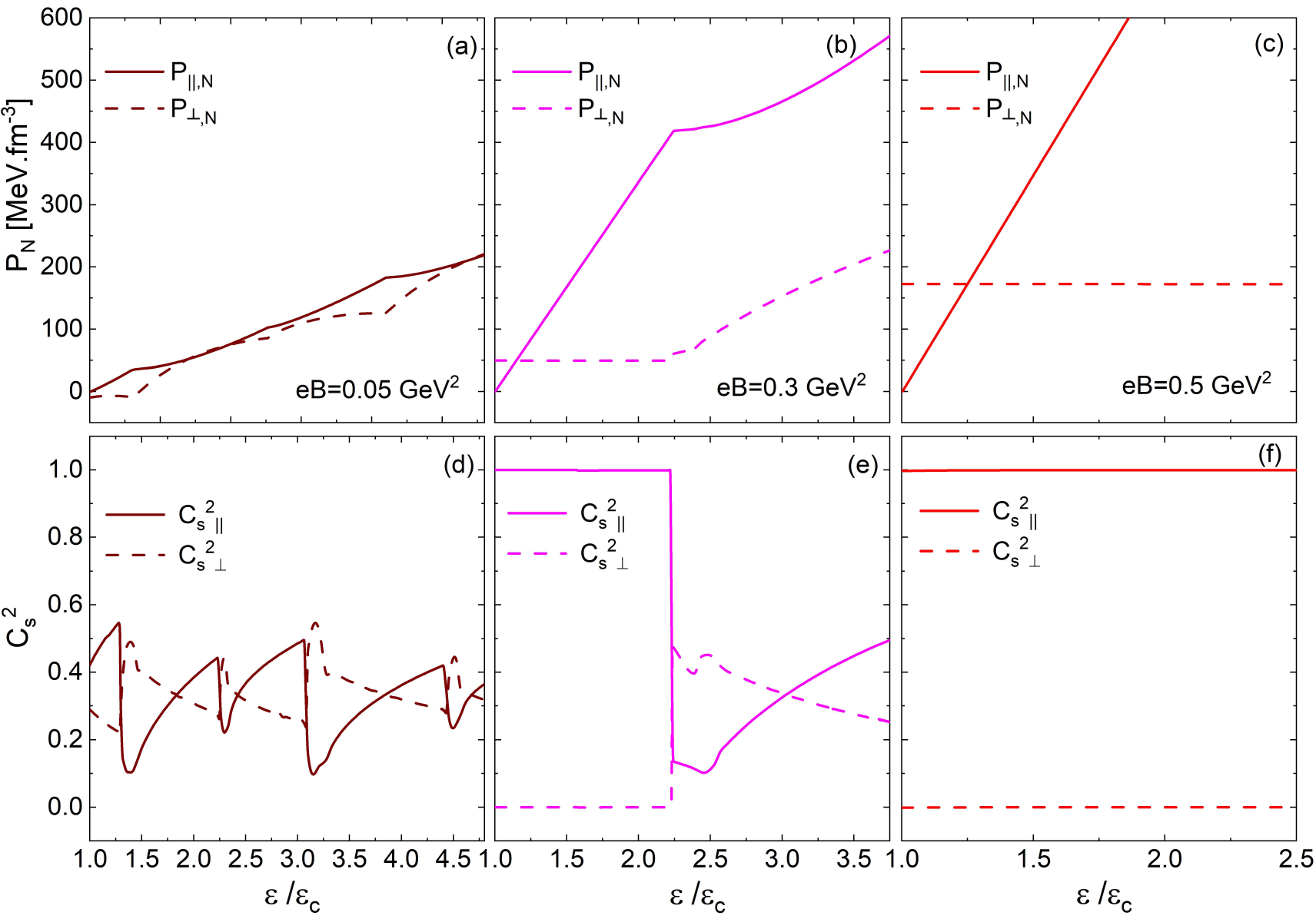}
\caption{Equation of state (upper panels) and squared speed of sound $C_s^2$ as a function of normalized energy density $\varepsilon/\varepsilon_c$ (lower panels) for the parallel (solid) and perpendicular (dashed) directions with respect to the magnetic field. Three different values for magnetic field are displayed: $eB=0.05$ GeV$^2$ (left, $\varepsilon_c=240.5$ MeV/fm$^3$),  $eB=0.3$ GeV$^2$ (center, $\varepsilon_c=337.3$ MeV/fm$^3$) and $eB=0.5$ GeV$^2$ (right, $\varepsilon_c=698.3$ MeV/fm$^3$)}
\label{fig:eos_and_Cs2_merge}
\end{figure} 

\end{widetext}

Figure~\ref{fig:magnetiz_merge}(a)–(b) shows that the normalized magnetization $\mathcal{M}_N$ is negligible while the system remains in the $\chi SB$ phase and becomes finite after the chiral transition, entering a nonmonotonic regime governed by Landau-level population. As a function of $eB$, Figure~\ref{fig:magnetiz_merge}(a), $\mathcal{M}_N$ develops de Haas–van Alphen–like oscillations at moderate fields that are gradually suppressed as the system approaches the LLL-dominated regime; as a function of $\mu$ (panel b), $\mathcal{M}_N(\mu)$ rises within each Landau interval and then drops when the next level opens. These features are present in both the chiral and finite-mass calculations and, at low fields, are in qualitative agreement with local-NJL results~\cite{Chaudhuri:2022oru}, while the nonlocal model presented here extends the range into the strong-field domain.
The magnetic susceptibility shown in Fig.~\ref{fig:suscept} provides a complementary, more oscillatory signature of the same physics: it alternates in sign at low $eB$ due to successive level thresholds, with the oscillation amplitude decreasing as $\mu$ grows, and it progressively approaches a positive limit in the LLL regime. In short, both $\mathcal{M}_N$ and $\chi_M$ display the Landau-quantization fingerprints already discussed for the chiral limit; here we only emphasize that these qualitative signatures persist for finite current masses.
Furthermore, the oscillatory behavior at low $eB$ and its suppression at higher fields still closely resemble the pattern reported in Fig.~\ref{fig:eos_and_Cs2_merge}(c) of Ref.~\cite{Chaudhuri:2022oru}, supporting the consistency of our findings with earlier effective-model studies

We conclude our analysis by presenting the equation of state and the squared speed of sound in the $\chi SR$ phase. These quantities provide direct insight into the stiffness of magnetized quark matter and the propagation of perturbations in the medium, complementing the pressure and magnetization results discussed previously. 

For finite current quark masses, Fig.~\ref{fig:eos_and_Cs2_merge} shows the pressures (a)–(c) and the squared speed of sound (d)–(f) for three representative values of the magnetic field. Their qualitative behavior closely follows the chiral limit presented in Fig.~\ref{Fig:EoS_Cs2_chiral}. The longitudinal pressure stiffens with increasing $eB$, whereas the transverse pressure is suppressed and may even become small or negative in the strong field regime. Similarly, $C_{s,\parallel}^2$ increases with both $\varepsilon$ and $eB$, approaching the causal bound in the LLL-dominated region, while $C_{s,\perp}^2$ decreases monotonically, reflecting the reduced transverse dynamics.

The main difference with respect to the chiral limit shown in Fig.~\ref{fig:density_merge_chiral} lies in the character of the transitions: the qualitative trends remain the same, but finite masses soften the discontinuities in thermodynamic observables. 
In the low-field regime, the pressures and the speed of sound closely follow the chiral case, with successive transitions associated with the filling of Landau levels. 
In contrast, at high magnetic fields the essential picture becomes robust and universal: the longitudinal response is strongly enhanced, the transverse one is suppressed, and the equation of state develops a pronounced anisotropy.

\section{Summary and conclusions}
\label{sect4}

In this work we have performed a detailed analysis of cold magnetized quark matter within the framework of a covariant nonlocal two-flavor Nambu–Jona-Lasinio (nlNJL) model, considering both the chiral limit and the case with finite current quark masses. We have studied several thermodynamic quantities—such as the anisotropic pressures $P_{\parallel}$ and $P_{\perp}$, the quark number density, magnetization, magnetic susceptibility, and the directional speed of sound—over a wide range of magnetic fields and chemical potentials. 

Although our analysis encompasses the transition from the chirally broken ($\chi SB$) to the chirally restored ($\chi SR$) phase, we focused mainly on the latter, where the discrete Landau-level structure and the resulting anisotropic effects become most pronounced. 
In the strong-field limit, the system enters an effective $(1+1)$-dimensional regime where the longitudinal dynamics dominate.

Overall, we find that both the chiral and finite-mass cases display qualitatively similar behavior, showing that the essential mechanisms driving the thermodynamic response and pressure anisotropy are robust against the inclusion of finite quark masses.

Our results show that the magnetic field induces a pronounced anisotropy between the longitudinal and transverse pressures, which increases with field strength. The difference $P_{\parallel} - P_{\perp}$ reflects the system’s magnetization and encodes the underlying Landau-level structure. The magnetic field enhances the quark density and stiffens the longitudinal equation of state, while the transverse pressure is partially reduced due to the magnetic contribution. The oscillations associated with successive Landau-level fillings are clearly observed in both cases, and their amplitude and pattern are comparable, indicating that finite current quark masses do not diminish the strength of these effects.

The analysis of the quark density $n$ further confirms the role of Landau quantization and magnetic catalysis. At low magnetic fields, $n(\mu)$ exhibits staircase-like structures corresponding to the successive population of higher Landau levels, while at strong fields the system becomes dominated by the lowest Landau level (LLL), resulting in a nearly linear increase of $n$ with $eB$. Finite quark masses preserve the qualitative trends, including the first-order nature of the transitions. Magnetic catalysis thus emerges as a robust feature of dense quark matter, largely insensitive to the details of the current quark mass.

The behavior of the magnetization shows that both the chiral and finite-mass cases exhibit the same qualitative pattern. In the $\chi SR$ phase, $\mathcal{M}_N$ increases with $eB$ and $\mu$, displaying oscillations associated with the sequential filling of Landau levels that gradually fade as the system enters the LLL-dominated regime. The field dependence obtained here closely follows that of the chiral limit, confirming that finite current quark masses do not alter the underlying mechanisms driving the magnetic response. At low fields, our results are consistent with those of Chaudhuri \textit{et al.}~\cite{Chaudhuri:2022oru}, while at higher $eB$ the nonlocal model naturally extends the behavior into the strong-field domain.

The equation of state reveals a progressive stiffening of the longitudinal sector as $eB$ increases, approaching the limit $P_{\parallel}\simeq \varepsilon$ characteristic of a $(1+1)$-dimensional system. Consequently, the squared speed of sound in the longitudinal direction tends to the causal bound, $C_{s,\parallel}^2 \to 1$, in the strong-field LLL regime. In contrast, the transverse sector exhibits a reduction of $P_{\perp}$ and a decrease in $C_{s,\perp}^2$, which can even become negative, signaling potential instabilities of transverse modes. These effects persist with finite quark masses, demonstrating that the effective dimensional reduction induced by strong magnetic fields is a universal phenomenon.

The magnetic susceptibility $\chi_M$ provides further insight into the system's response. At low magnetic fields, $\chi_M$ oscillates around zero, whereas at higher fields, the oscillations disappear and the susceptibility stabilizes at positive values. This transition from an oscillatory to a stable regime is consistent with the behavior of the pressures and density, highlighting the dominant role of the LLL in shaping the system's properties.

We have also compared our results with those obtained using the local NJL model of Ref.~\cite{Chaudhuri:2022oru}, where finite current quark masses are considered and magnetic fields are limited to $eB \lesssim 0.16~\mathrm{GeV}^2$. Within this range, both our nonlocal calculations—with and without finite current quark masses—exhibit the same qualitative trends found in the local model, including the behavior of the normalized longitudinal and transverse pressures, as well as of the magnetization and magnetic susceptibility. This agreement supports the robustness of the physical mechanisms driving the anisotropic behavior of magnetized quark matter across different effective approaches.

As a future line of research, we plan to extend this study to systems at finite temperature and vanishing chemical potential, in order to investigate the temperature dependence of the magnetic-field-induced anisotropy. This analysis will enable a direct comparison with available lattice QCD results. In addition, it would be of interest to explore charge-neutral matter under $\beta$-equilibrium conditions, which are relevant for the description of dense stellar environments such as neutron stars~\cite{Patra:2020wjy,Becerra:2024wku}. The results of these projects are expected to be reported in the near future.
 
\begin{acknowledgments}
The authors would like to acknowledge the financial support from CONICET under Grant No. PIP 22-24 11220210100150CO, ANPCyT (Argentina) under Grant PICT20-01847, and the National University of La Plata (Argentina), Project No. X960.

\end{acknowledgments}

\bibliography{TMUB}

\end{document}